\documentclass[iop,apj,tighten]{emulateapj}

\slugcomment{To appear in the Astrophysical Journal.}

\shorttitle{Deep structure of solar meridional flow}
\shortauthors{Rajaguru and Antia}

\begin{document}


\title{Meridional circulation in the solar convection zone: time-distance helioseismic inferences from four years of HMI/SDO observations}


\author{S.P. Rajaguru}
\affil{Indian Institute of Astrophysics, Koramangala II Block,
    Bangalore 560034, India.}
%\email{rajaguru@iiap.res.in}
%\altaffiltext{*}{\email{rajaguru@iiap.res.in}}
\and

\author{H.M. Antia}
\affil{Tata Institute of Fundamental Research, Homi Bhabha Road, Mumbai 400005, India}



\begin{abstract}

We present and discuss results from time-distance helioseismic measurements of meridional circulation in the solar convection zone 
using 4 years of Doppler velocity observations by the Helioseismic and Magnetic Imager (HMI) onboard the Solar Dynamics Observatory 
(SDO). Using an in-built mass conservation constraint in 
terms of the stream function we invert helioseismic travel times to infer meridional circulation in the solar convection zone. 
We find that the return flow that closes the meridional circulation is possibly beneath the depth of $0.77 R_{\odot}$. We discuss the 
significance of this result in relation to other helioseismic inferences published recently and possible reasons for the differences 
in the results. Our results show clearly the pitfalls involved in the measurements of material flows in the deep solar interior given 
the current limits on signal-to-noise and our limited understanding of systematics in the data. We also discuss the implications 
of our results for the dynamics of solar interior and popular solar dynamo models.

\end{abstract}



\section{Introduction}

The structure and dynamics of large-scale material flows on the solar surface and in the convection zone play fundamental roles in the 
working of solar dynamo and its cyclic variation \citep{pcharb10}. Circulating currents in planes through the meridians inside a 
rotating star were originally proposed by \citet{eddington25} as a consequence of radiative equilibrium \citep{vonzeipel24} and as 
those responsible for the development of differential rotation. A rigorous theoretical treatment of the origin and maintenance of 
meridional circulation (MC) requires dynamical models that include energy and momentum transfer between convection, differential 
rotation and thermal stratification \citep{gilmanmiller86,mieschtoomre09}. Mean-field hydrodynamical models of global
axisymmetric flows and heat transport have been studied \citep{kitchatinovrudiger99,kitchatinovolemskoy11}; these models calculate interior
differential rotation and MC. However,  models with sufficient realism so as to reliably 
predict the deep structure of MC, with constraints from the helioseismically well determined solar interior rotation, are not yet 
established, although, detailed studies are in the offing (see \citet{feathmiesch15} and references therein). 

Observational studies, however, have led to considerable details on the velocity amplitudes (typically of 10 -- 20 m s$^{-1}$) and 
variation of surface and near-surface part of MC \citep{kommetal93,hathaway96,haberetal02,ulrich10,hathawayrightmire10}, and on its 
contribution to the transport of magnetic flux from lower latitudes to the poles \citep{wangetal89}. This latter aspect of MC, with a 
modeled mass-conserving amplitude for a return flow typically near the base of the convection zone, has been a key dynamical component 
in the flux transport dynamo models \citep{choudhurietal95,dikpaticharb99}. Precisely for the above reasons, reliable observational 
(helioseismic) inferences of the deep structure of MC is of paramount importance for understanding the origin and drivers of solar 
variability on cyclic time-scales.

Meridional flows on the surface have been measured using several different techniques: directly from the Doppler observations of the 
surface motions \citep{hathaway96,ulrich10}, correlation tracking of surface features such as small-scale magnetic elements 
\citep{kommetal93,hathawayrightmire10}, and local helioseismic measurements 
\citep{gilesetal97,schoubogart98,basuetal99,haberetal00,haberetal02, gonzalezetal08,basuantia10}. Almost all these different 
measurements give typical amplitudes of 10 -- 20 m s$^{-1}$ for the surface poleward flow with peak values over the latitude range of 30 -- 
50$^{\circ}$. Of the above only the helioseismic analyses have provided maps of flows in the sub-surface layers. While the ring 
diagram analyses cover sub-surface layers to a maximum depth of about 20 Mm, the time-distance helioseismic technique 
\citep{duvalletal93} can cover the flow in deeper layers. A first attempt application of time-distance helioseismology 
\citep{gilesetal97} to measure MC using the SOHO/MDI Doppler observations indicated that the poleward flows extended from the surface 
down to about 27 Mm with a roughly constant magnitude. Since then, while the ring diagram analyses have studied in detail the MC in 
the near-surface layers including its temporal (solar-cycle scale) variations \citep{gonzalezetal08,basuantia10}, a recent major 
development in the application of time-distance helioseismology has been the identification and removal of a large systematic 
center-to-limb effect (cf. Section 2) in the measured travel times of acoustic waves \citep{zhaoetal12,zhaoetal13}. This has led to 
possibility of studying MC using time-distance helioseismology throughout the convection zone, especially improving the identification 
of flow signals in the deeper layers \citep{zhaoetal13,jackiewiczetal15}. A new feature in these recent results, obtained from 
SDO/HMI data, has been the inference of a relatively shallow return flow starting at about 0.9R$_{\odot}$ and a possible second cell 
of MC below this depth \citep{zhaoetal13}. While the later study by \citet{jackiewiczetal15}, using data from the Global Oscillation 
Network Group (GONG) observations, agrees on the shallow return flow, it has not reproduced the deeper second cell of MC. Both the 
above studies have used about two years of data from SDO/HMI or GONG Doppler observations. Given the high stakes that these results 
have on the dynamics of solar interior and the working of a large class of solar dynamo models, it is important to check the 
consistency and reproducibility of these results. Both of the above studies have used mass-conservation constraints on the inverted 
flows for checking the physical validity of results. Here, we improve on these results as follows: (1) we use 4 years of SDO/HMI to 
improve on the signal-to-noise of travel time measurements, and (2) implement an in-built mass conservation constraint in terms of 
stream functions in the inversion process itself, thereby consistently deriving both the horizontal (latitudinal) and radial 
flows.

Although, to first order, global mode frequencies are not sensitive to meridional flow, application of quasi degenerate perturbation 
theory \citep{lavely92} shows that modes with close frequencies for neighbouring values of degree, $l$ can be coupled and their 
frequencies can be affected by meridional flows. \citet{schadetal13} have used the resulting perturbation on the eigenfunctions to 
study the MC in solar convection zone. Using MDI data for 2004--2010 they find evidence that MC penetrates to the base of the 
convection zone and the flow profile shows multiple cells in latitude and depth. This pattern is quite different from those inferred 
using time-distance technique and such a pattern is not seen in near surface flows which are reliably determined using the time 
distance and ring diagram techniques. The reason for this discrepancy is not clear, but it could be due to sensitivity of global mode 
perturbation to MC with multiple cells in latitude, as for multiple cells the probability of finding mode pairs (which are coupled by 
MC) with close frequencies increases.

Another technique to estimate MC is to use well determined differential rotation profile in the solar convection zone and solve a 
hydrodynamic model including the Coriolis force and Reynolds stresses \citep{dikpati14}. For a solar like density profile and 
differential rotation, the resulting meridional flow was found to be two-celled in latitude. Such a profile is not seen at the 
surface, which implies that other effects need to be included in such an analysis.

We describe the data and travel-time measurement procedure in Section 2, inversion technique in Section 3, present the results and 
validation tests on MC in Section 4 and discussions and conclusions in Section 5.

\section{Data and Analysis Procedure}

\subsection{Data}

We have used the full-disk Doppler observations made by the SDO/HMI over a four year period from 2010 May 1 through 2014 April 30. The 
data consists of 45 s cadence Doppler velocities at the solar surface. The first 2 years of data are the same as those used by 
\citet{zhaoetal13}, and we have added the next 2 years' data to it. The data are tracked to account for the solar rotation and 
remapped using Postel's projection to have a uniform latitude and longitude grid. This process is the same as that performed by 
\citet{zhaoetal13} using the JSOC helioseismology pipeline at Stanford. The original extracts from the JSOC database for this study 
used a spatial sampling scale of 0.18 deg/pixel, and we further resample (or smooth) the data to a scale of 0.36 deg/pixel in order to 
reduce the volume of data handled and the computing time. This increase in pixel size (or reduction in resolution) does not make any 
difference to the travel time measurements over distances greater than 2 deg., and further our main focus is on deeper layers which 
are probed by waves skipping at the surface over distances much larger than a degree.

\subsection{Analysis Procedure}

We use the technique of time-distance helioseismology \citep{duvalletal93} in the so called deep-focus geometry, in much the same way 
as that described by \citet{zhaoetal13} except for implementing an improvement as discussed below. This method involves 
cross-correlating Doppler signals from arcs, typically 30 degree wide, cut out from an annulus of diameter equaling the surface travel 
distance of the helioseismic waves, whose travel times that we want to measure. These arcs are placed perpendicular to the direction 
over which we want to measure the flows ({\it e.g.,} for meridional flows the arcs are placed perpendicular to the meridians). 
Ray-paths connecting diametrically opposite points in the arcs meet at the lower turning point of waves traveling a surface distance 
of $\Delta$ that equals the diameter, and it is also the deep-focus location. This location is directly beneath the centre point of 
the arcs (or annulus) at the surface, hence the measurement is assigned to this surface point. Measurements are repeated by moving the 
centre point to every pixel of the area that we want to cover on the solar surface. For each measurement, \citet{zhaoetal13} took the 
average of Doppler signals over all pixels in one arc and cross-correlated it with that from the opposite arc. This averaging 
procedure, however, does not preserve the travel distance ($\Delta$ = the diameter) as it involves contributions from all pairs of 
points (or pixels) between the arcs not just the diametrically opposite ones. In other words, a travel-time measurement by 
\citet{zhaoetal13} for a given $\Delta$ involves contributions from waves of smaller distances down to $\approx$ 
cos(15$^{\circ})\Delta$ {\footnote{This estimate holds when neglecting the spherical curvature on the solar surface, i.e. when the 
distances $\Delta$ are not large. Otherwise, exact expression for the smallest distance, $\alpha$, between pairs of points from 
opposite arcs cut from a circle of diameter $\Delta$ is given by $\alpha$ = 2 sin$^{-1}$[sin($\Delta$/2)cos($\theta$/2)], where 
$\theta$ is the arc-length in degrees}}, and hence measurements are biased towards lower depths at each depth, and 
this we believe will lead to poorer depth resolution as $\Delta$ increases, i.e. for deeper measurements. We correct for this in our 
measurements, but at the cost of much increased computing time, by splitting the arc into small segments (typically 1$^{\circ}$ 
to 3$^{\circ}$ segments, depending on $\Delta$) and cross-correlating the diametrically opposite segments [such a deep-focus geometry 
was originally proposed by \citet{duvall95}; see also \citet{rajaguru08,hanasogeetal10}], and then averaging the resulting 
cross-correlations for a given measurement. We denote travel times estimated from such 'point-to-point' cross-correlations with a 
string $p-p$. We also calculate cross-correlations between arc-averaged signals, exactly replicating the procedure of 
\citet{zhaoetal13}, and the travel times from such 'arc-to-arc' cross-correlations are denoted as $a-a$ (refer to Figure 2).

The rest of our measurement procedure is identical to \citet{zhaoetal13} in making one travel time measurement for each month (of four 
years of data that we use here): (1) a cross-correlation computation at a given location involves full-disk data cubes tracked for one 
day, (2) measurement is restricted to $30^{\circ}$ (in heliographic coordinates) wide strips centered about the central-meridian (for 
flows in the North - South direction) or the equator (for flows in the West - East direction) [the W-E travel time differences, after 
accounting for rotation signals, capture the centre-to-limb systematics in travel times that are to be subtracted from the N-S travel 
times \citep{zhaoetal12,zhaoetal13}, see below], (3) each day's cross-correlation functions are averaged over the same latitudes 
(longitudes) for N-S (W-E), and then averaged again over one month intervals, and (4) each monthly averaged cross-correlation function 
is fitted by a Gabor wavelet \citep{sashaduvall97} to estimate the acoustic wave travel times at each latitude (for N--S) or longitude 
(for W--E) in two opposite directions between the arcs. The difference between the travel times in the two opposite directions is 
related to the flow velocities in the region traversed by the waves \citep{duvalletal93,gilesetal97}. For example, for the N--S 
aligned arcs, the difference $\delta\tau_{\rm{NS}} = \tau_{\rm{NS}} - \tau_{\rm{SN}}$ between the N--S and S--N travel times will be 
positive for a flow towards the north pole. We have employed 60 travel distances, $\Delta$, ranging between $2.16^{\circ}$ and 
$44.64^{\circ}$ in steps of $0.72^{\circ}$, covering a depth range from near the surface down to about $0.7 R_{\odot}$, again the same 
as that of \citet{zhaoetal13}.

As shown by \citet{zhaoetal12}, there is a large systematic increase in travel time differences against angular distance from the 
solar disk center mimicking a radial outflow from the centre towards the limb, and which increases as $\Delta$ increases. This large 
center-to-limb systematics (CLS) in travel times is still of unknown origin, although, the analyses of \citet{zhaoetal12,zhaoetal13} 
involving comparisons of travel times from different observables (corresponding to different heights of formation in the solar 
atmosphere) from the SDO/HMI as well as with that from another instrument (SOHO/MDI) pointed to possible physical causes in the solar 
atmosphere related to observation height differences. A study by \citet{baldnerschou12} showed that the near-surface granular 
convection could affect the wave-propagation in the observable layers leading to a similar effect as the CLS in travel times. Although,
a consistent explanation of the origin of CLS in helioseismic measurements from different observables is yet to be achieved, it is 
clear that the CLS in travel times are not related to sub-surface flows. The empirical prescription to remove them in the travel 
times, as demonstrated by \citet{zhaoetal12}, has led to improved measurements of meridional flows.

\section{Travel-time Inversion Technique}

Within the ray approximation,
the travel-time difference $\delta\tau$ between the two opposite directions along a ray path $\Gamma_0$ is given
by \citep{sasha96}:
\begin{equation}
\delta\tau=-2\int_{\Gamma_0} \frac{\mathbf{u}\cdot{\hat\mathbf{n}}}{c^2}\;ds,
\label{eq:inv}
\end{equation}
where $\mathbf{u}$ is the flow velocity and $\hat\mathbf{n}$ is the unit vector
along the ray path. For ray paths confined to N--S plane, only the meridional
flow will contribute to the travel-time. In general the contribution from
the $u_\theta$ component would be much larger than that from the $u_r$ component
of the meridional flow.  This is because in general, the $u_\theta$ component
is larger than $u_r$ and further contribution from $u_r$ tends to cancel out
between the rising and falling branch of the ray path, while $u_\theta$
contribution in the two branches add to each other.

Because of small contribution from $u_r$ to $\delta\tau$, it is not possible
to directly determine this component from the measured travel-times. However,
the two components of the meridional flow are not independent, as they satisfy
the continuity equation. Thus it is possible to determine both components
and to satisfy the continuity equation, if instead we determine the stream
function, $\psi$, which gives:
\begin{eqnarray}
\rho u_r&=&\frac{1}{r}\frac{\partial\psi}{\partial\theta}+
\frac{\cos\theta}{r\sin\theta}\psi,\\
\rho u_\theta&=&-\frac{\partial\psi}{\partial r}-\frac{\psi}{r},
\end{eqnarray}
where $\rho$ is the density in a solar model that is used to calculate the
ray paths. Thus we use Eq.~\ref{eq:inv} to determine $\psi$ using the
observed travel-times. Because of large variation in density, $\psi$ also
shows similar variation, as a result we actually use $\psi'=\psi/\rho$
for inversions.

We use the Regularized Least Squares (RLS) technique to solve the inverse
problem to calculate $\psi'$ in the convection zone. For this we expand
$\psi'(r,\theta)$ in terms of cubic B-splines (e.g., \citet{antia12}) in $r$ and $\theta$:
\begin{equation}
\psi'(r,\theta)=\sum_i\sum_j a_{ij}\Phi_i^r(r)\Phi_j^\theta(\theta),
\end{equation}
where $\Phi_i^r(r)$ are the cubic B-spline basis covering $0.69R_\odot\le r
\le R_\odot$
and $\Phi_j^\theta(\theta)$ are the cubic B-spline basis
covering $|\theta-\pi/2|\le 1.055$.
We use 38 knots in $r$ which are uniformly spaced in acoustic depth and
31 knots in $\theta$ which are uniformly spaced in $\theta$ to define the
B-spline basis functions. The coefficients
of expansion $a_{ij}$ are determined using RLS with second derivative smoothing
\citep{gough91}
in both $r$ and $\theta$ by minimizing the function
\begin{equation}
\sum_i\left(\frac{d_i}{\sigma_i}\right)^2+\lambda_r^2\sum\left(\frac{\partial^2
\psi'}{\partial r^2}\right)^2+\lambda_\theta^2\sum\left(\frac{\partial^2
\psi'}{\partial \theta^2}\right)^2,
\label{eq:rls}
\end{equation}
where $d_i$ are the residual in the fit to Eq.~\ref{eq:inv} (i.e., the
difference between the left hand side and right hand sides of the equation) and
$\sigma_i$ are the corresponding errors in travel time differences.
Here the summation in the last two terms are over all knots in the B-spline
representation and $\lambda_r$ and $\lambda_\theta$ are the two smoothing
parameters.
Apart from smoothing we also apply the boundary condition $\psi'=0$ at
the upper boundary. In principle, the same condition should be applied at the
lower boundary, if the flow is contained within the volume considered. Since
we do not know the lower boundary where the flow ends, we have not applied this
boundary condition. We have verified that adding this boundary condition forces
$u_r$ to vanish at the lower boundary ($r=0.69R_\odot$) and the resulting solution
is modified in the neighbourhood of the boundary, but the solution over the
bulk of the region is unaffected.
These smoothing parameters are chosen as the
minimum values that are needed to obtain
a solution that is smooth. Once these coefficients
are known, $u_r$ and $u_\theta$ are easily computed over the entire interval.
Because of very small density scale height in the near surface region, we
encountered some difficulty in calculating $\psi'$ in this region.
This problem is probably because our knot spacing in this region is much
larger than the density scale height.
To avoid
this we use the value of $\psi'$ at $r=0.995R_\odot$ for larger values of $r$ also.
This may be justified as earlier studies do not show much variation in
the meridional flow velocity above this layer. Further, we find that removing
this artifact only affects the solution for $r>0.99R_\odot$, which is not
the region where we are primarily interested in this work.

To calculate the errors in inverted flow velocities, we repeat the
calculations 100 times with $\delta\tau$ randomly perturbed with estimated
errors in observed values. In addition to estimate the systematic error due
to choice of smoothing parameters we also perturb these randomly. The standard
deviation in these values would give an estimate of errors in inversion.

\section{Results}
\subsection{Travel-time Differences due to Meridional Flows}

Travel-times of waves skipping at distances, $\Delta$, on the surface ranging between $2.16^{\circ}$ and $44.64^{\circ}$ correspond to 
lower turning points, $r_{t}$, between 0.987R$_{\odot}$ and 0.701R$_{\odot}$, thus covering the whole depth range of the convection 
zone. Hence, we expect that the travel time differences $\delta\tau_{\rm{NS}}$ that result after subtracting the CLS (estimated from 
$\delta\tau_{\rm{WE}}$ after correcting for solar rotation) for the chosen range of $\Delta$ should have signatures of the MC in the 
whole of convection zone. We show $\delta\tau_{\rm{NS}}$ for a few selected $\Delta$ in Figure 1. These travel times are estimated 
from the $p-p$ measurement scheme described in Section 2.2. As described earlier, each measurement shown in Figure 1 is an average of 
individual measurements that span the given small range of $\Delta$ (indicated in the panels) with a spacing $0.72$ deg. and a time 
range of 47 months (in the four year period 2010 May 1 through 2014 April 30, we have the month of April 2012 missing in our data 
sets, and hence we have a total of 47 months) with one measurement for each month. The error bars represent standard errors estimated 
from these individual measurements. The errors in $\delta\tau_{\rm{NS}}$ and $\delta\tau_{\rm{WE}}$ are of similar magnitude. We have 
also removed a small offset in $\delta\tau_{\rm{NS}}$ resulting from the $P$ angle variation, i.e. the variation of the angle the 
solar rotation axis makes with respect to the instrument (spacecraft), following \citet{gilesetal97} and \citet{zhaoetal13}.

\begin{figure}
\epsscale{1.2}
\plotone{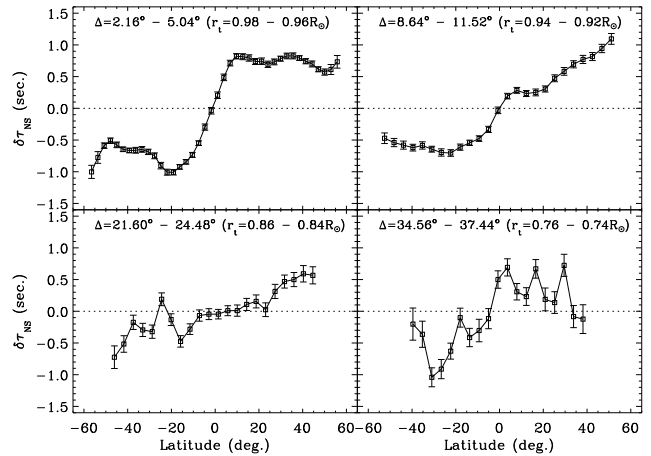}
\caption{Travel-time differences, $\delta\tau_{\rm{\rm{NS}}}$, against latitude for four selected travel distances ($\Delta$). Each measurement shown here is an average of 
individual measurements, with a spacing $0.72$ deg. in $\Delta$ and one for each of 47 months. The error bars represent
standard errors estimated from these individual measurements. The center-to-limb systematics in travel times estimated through $\delta\tau_{\rm{WE}}$ have been subtracted from $\delta\tau_{\rm{NS}}$.}
%See the electronic edition of the Journal for a color version
%of this figure.
\label{fig1}
\end{figure}

Variation of $\delta\tau_{\rm{NS}}$, averaged over three different latitude ranges, against $\Delta$ (and hence against lower turning 
points of p modes) is shown in Figure 2. We have shown measurements from both our improved $p-p$ measurement scheme and the $a-a$ 
scheme of \citet{zhaoetal13}. In general, we find that travel times from both these schemes agree well for the few smaller $\Delta$, 
but there are significant differences for $\Delta > 30^{\circ}$ over low and mid-latitudes. For further analyses and inversions we use 
only the $p-p$ travel times. We believe that the variation seen against the three latitude ranges for $\Delta > 22^{\circ}$, and the 
North - South asymmetry in them capture the signatures of deep meridional circulation. It is interesting to note the differences in 
travel times against the latitude ranges used in the three panels of Figure 2, especially for large $\Delta$: travel time averages 
over $30^{\circ} - 40^{\circ}$ latitudes exhibit a roughly constant value of about 0.5$s$ over $15^{\circ} < \Delta < 32^{\circ}\; 
(0.9R_{\odot} > r_{t} > 0.8R_{\odot})$ and relatively large variations for $\Delta > 32^{\circ}$ (i.e. for $r_{t} < 0.8R_{\odot})$, 
especially a sharp decline (at $\Delta \approx 30^{\circ}$) and a rise again ($\Delta > 40^{\circ}$) in the northern hemisphere, 
although, the errors are relatively large here.
%Such a variation, we believe, is due to a deeper extent of poleward flows at these higher latitudes and is indicative of return flow at depths below 0.8R$_{\odot}$. 
%In contrast, the lower latitude signals for $\Delta > 30^{\circ}$ could likely be due to a more complicated structure than
%that of a single cell MC at deeper layers. Firm inferences on these can only be obtained from travel-time inversions, which we present in the following sections.}
\begin{figure}
\epsscale{1.65}
%\epsscale{.95}
\plotone{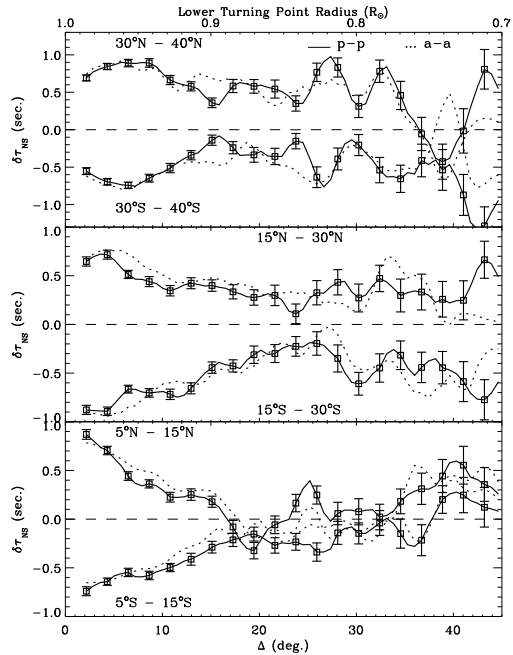}
%\plotone{td_lat_avg_multi5t15_15t30_30t40_nn.eps}
\caption{Travel-time differences, $\delta\tau_{\rm{NS}}$, averaged over three different latitude ranges, as marked in the panels, against the travel distance $\Delta$ and lower
turning point radius.}
\label{fig2}
\end{figure}

\subsection{Meridional Circulation from Inversions of Travel-times} 

As presented in Section 3, we have implemented a ray-theoretic scheme of inverting the travel time differences $\delta\tau_{\rm{NS}}$ 
to determine the stream function $\psi$, which allows the determination of both the flow components, $u_{\theta}$ and $u_{r}$, while 
satisfying the mass-conservation constraint. To illustrate the effect of choice of smoothing parameters we show the results 
obtained using two different values. The errors in the inverted results depend on the smoothing, with generally a higher smoothing 
leading to smaller errors. Results of thus determined $u_{\theta}$ and $u_{r}$ covering the whole of the convection zone in depth and 
within $\pm60^{\circ}$ latitudes are shown in Figures 3 and 4. The decreasing latitude extent for deeper layers are due to decreasing 
usable surface coverage of data for larger $\Delta$, $\it{i.e.}$ we restrict measurements within $\pm 60^{\circ}$ latitudes to avoid 
errors due to projection effects and this means that usable coverage reduces as $\Delta$ increases.

\begin{figure*}[ht]
\centering
\epsscale{1.17}
%\plottwo{uth2_ur2_polar_corrected_4y_n.eps}{uth2_ur2_vs_r_minerr1_n.eps}
\plottwo{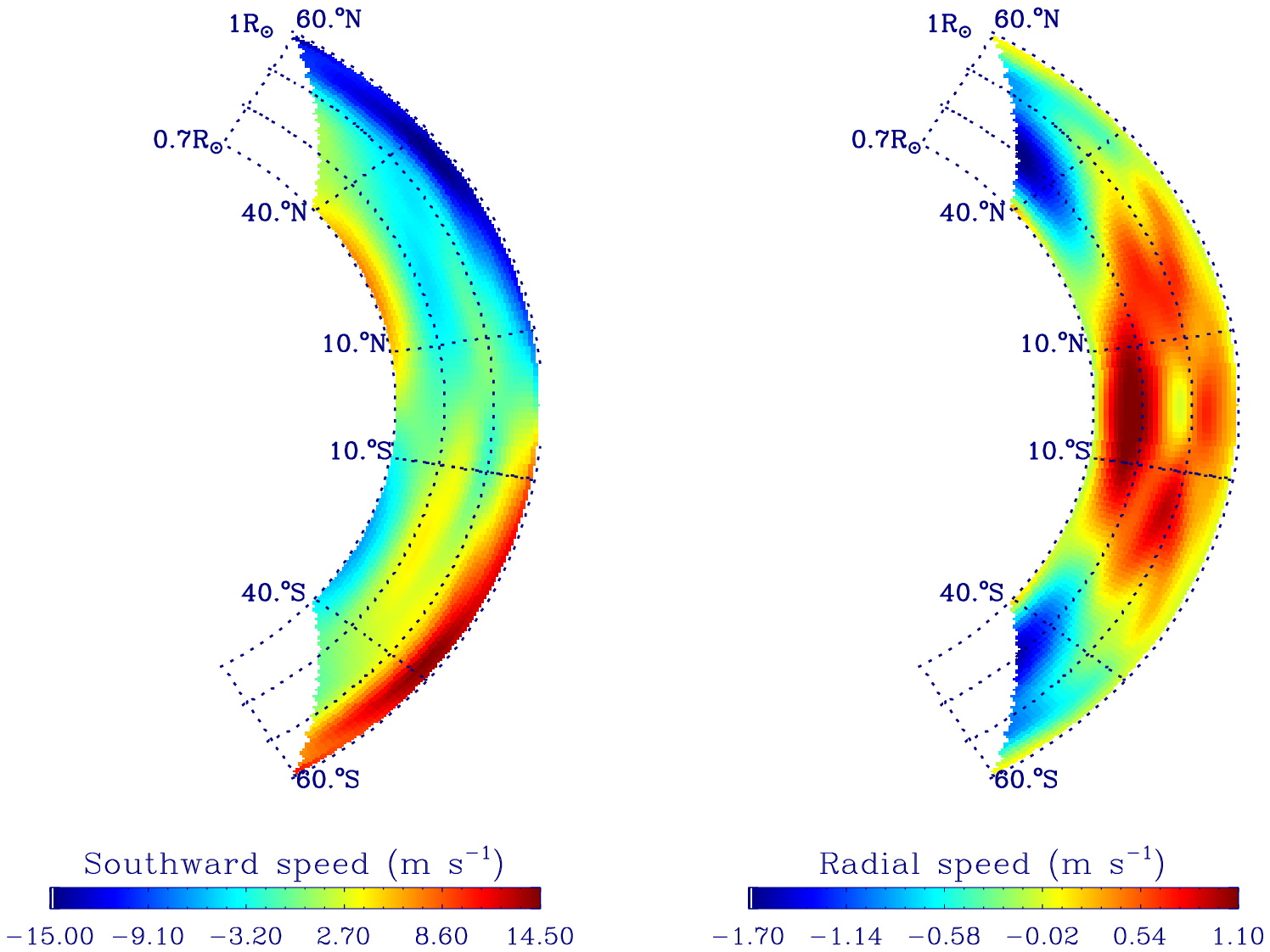}{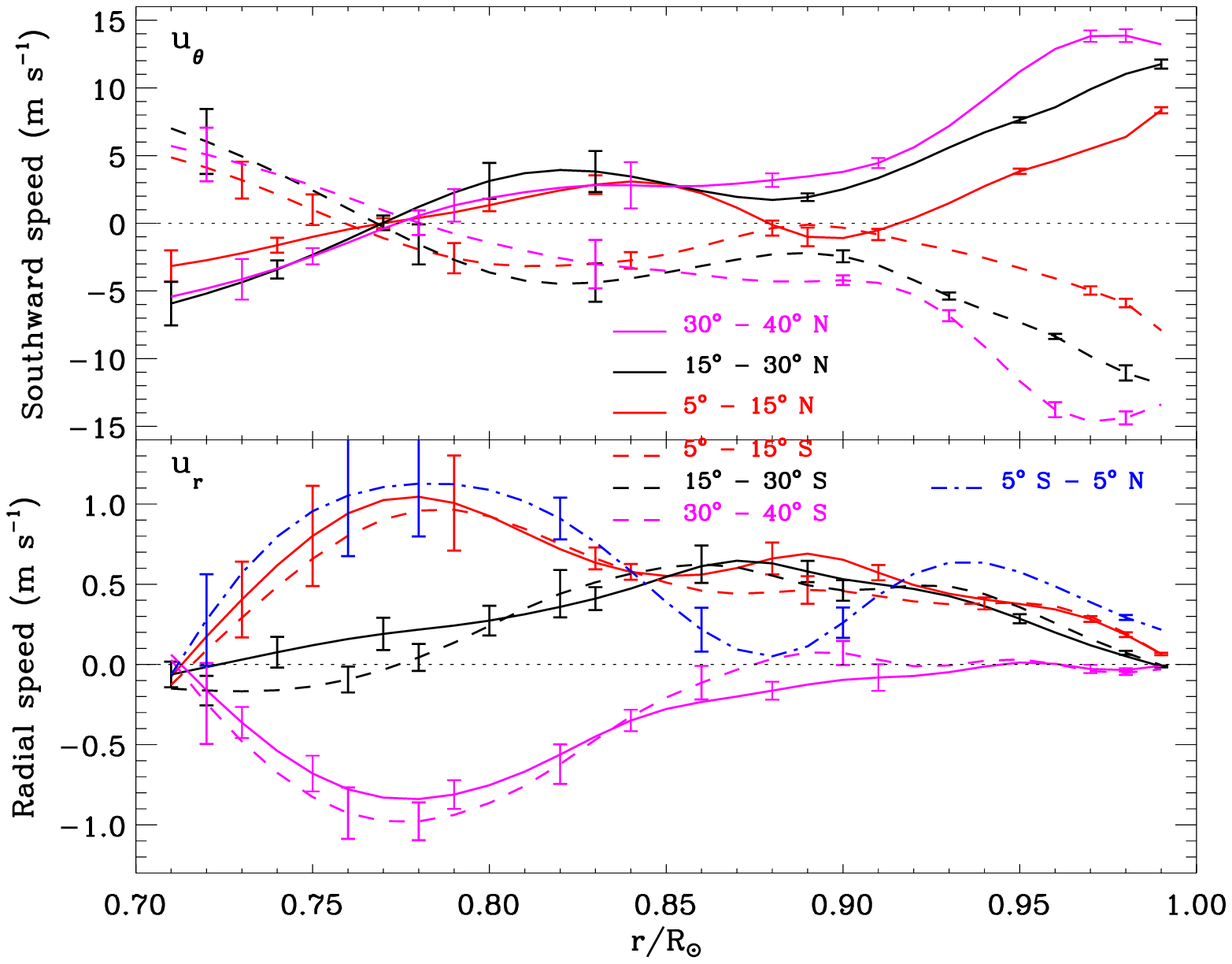}
\caption{Meridional circulation that results from inversions of travel-time differences, $\delta\tau_{\rm{NS}}$ for the case of higher smoothing
($\lambda_r=2\times10^{-4}, \lambda_\theta=5\times10^{-3}$).
The left part of the figure shows the 2D ($r$,$\theta$) profiles of $u_{\theta}$ and $u_{r}$. The two-panel plot on the right shows depth-profiles of $u_{\theta}$ and $u_{r}$ 
averaged over three different latitude ranges, as marked in the panel.}
\label{fig3}
\end{figure*}

Variations over depth of inverted $u_{\theta}$ and $u_{r}$, averaged over the same ranges of latitudes used in Figure 2 for 
$\delta\tau_{\rm{NS}}$, are shown in right panels of Figures 3 and 4. The lower smoothing lead to a slight increase in the variations 
in velocity profiles. These results show that the global large-scale MC has a possible return flow at depths below about $0.77 
R_{\odot}$. This large single-cell MC is clear at higher latitudes. However, at lower latitudes (less than about $30^{\circ}$) there 
appear to be some sign change in $u_\theta$ around $0.9R_\odot$, but the values are within error bars and these are also consistent 
with one deep cell of MC. Such variations look a little more pronounced in the case of solution with lower smoothing, as a comparison 
of Figures 3 and 4 would show. It can be seen that the results with two different smoothing shown in Figures 3 and 4 are not very 
different, thus showing that the solution is not particularly sensitive to choice of smoothing parameters in this range. Here we have 
only changed $\lambda_\theta$, but change in $\lambda_r$ also has similar effect. Nevertheless, below about $0.9R_\odot$ the magnitude 
of velocity is comparable to errorbars and hence we cannot rule out the possibility of multiple cells, especially at lower latitudes. 
The depth variation of radial component $u_{r}$ too (see right panels of Figures 3 and 4), near the equatorial region, seem to 
indicate a tendency of sign change at about $0.9R_\odot$, but again such changes are not significant considering the errors there. 
Zhao et al.~(2013) found return flow below about $0.9R_\odot$ and a second cell lower down. Considering the errorbars in inversion 
results, their results are roughly consistent with ours, though interpretation in terms of multiple cells is different. Although, the 
most recent study by \citet{jackiewiczetal15} using two years of data from GONG largely reproduces the shallow return flow below about 
$0.9R_\odot$, it does not show the second cell of MC deeper down and instead shows strikingly anti-correlated flows in deeper layers 
between GONG and HMI results. It is important to note that the amplitudes of $u_{\theta}$, in general, gradually decrease below about 
$0.97R_\odot$ in all results. Consistent inclusion of mass-conservation in our inversion process itself, is different from the results 
of both \citet{zhaoetal13} and \citet{jackiewiczetal15}. We believe this constitutes improved physical realism in our results than 
those of the above authors. We defer further discussion on the results to Section 5.
%However, it is interesting to note that the theoretical and simulation
%models of Featherstone and Miesch (2015) do predict a lower latitude multi-cellular structure for MC for solar-like rotation.
\begin{figure*}[ht]
\centering
\epsscale{1.17}
%\plottwo{uth2_ur2_polar_corrected_4ya.eps}{uth2_ur2_vs_r_minerr1_4ya_n.eps}
\plottwo{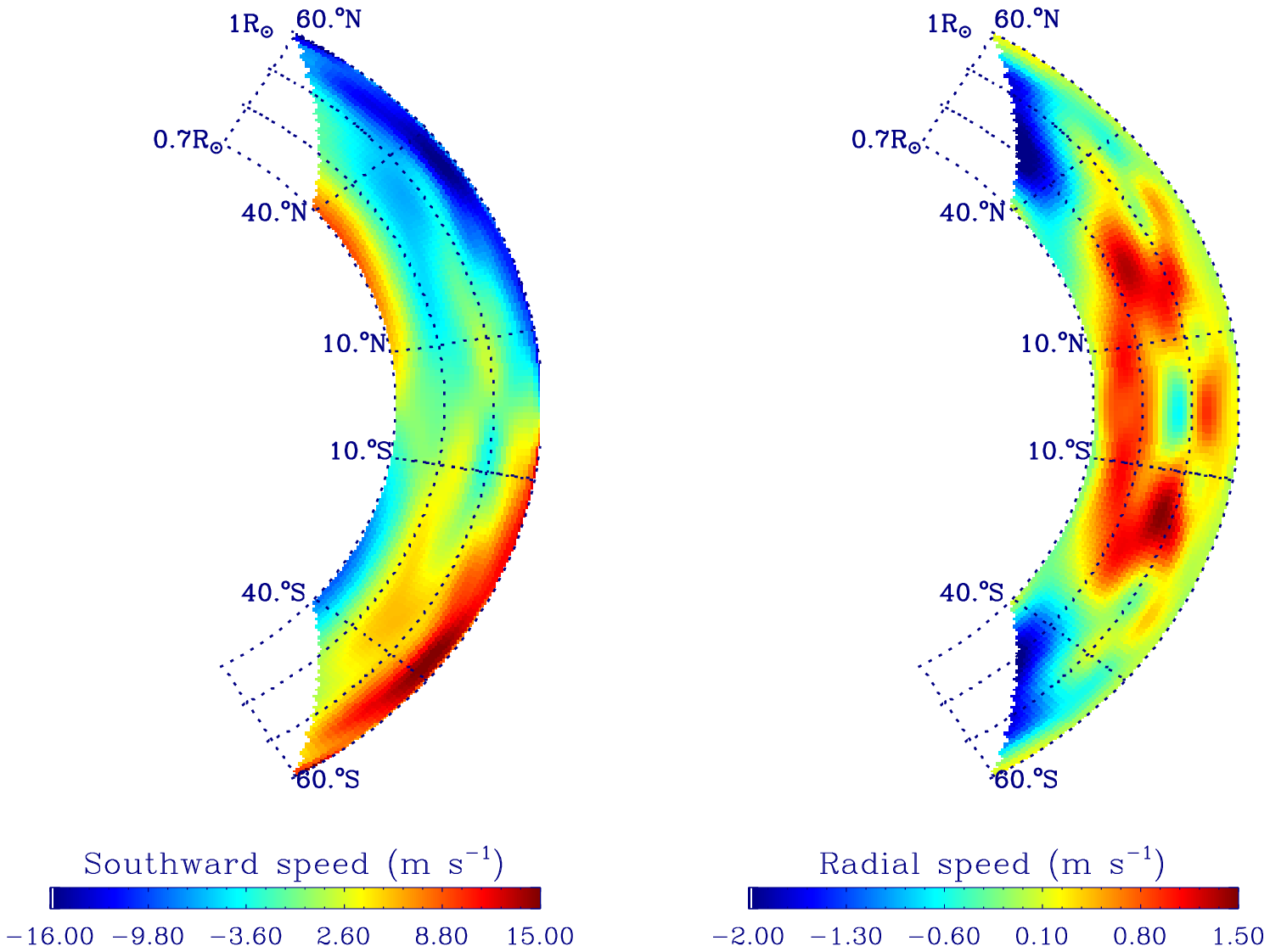}{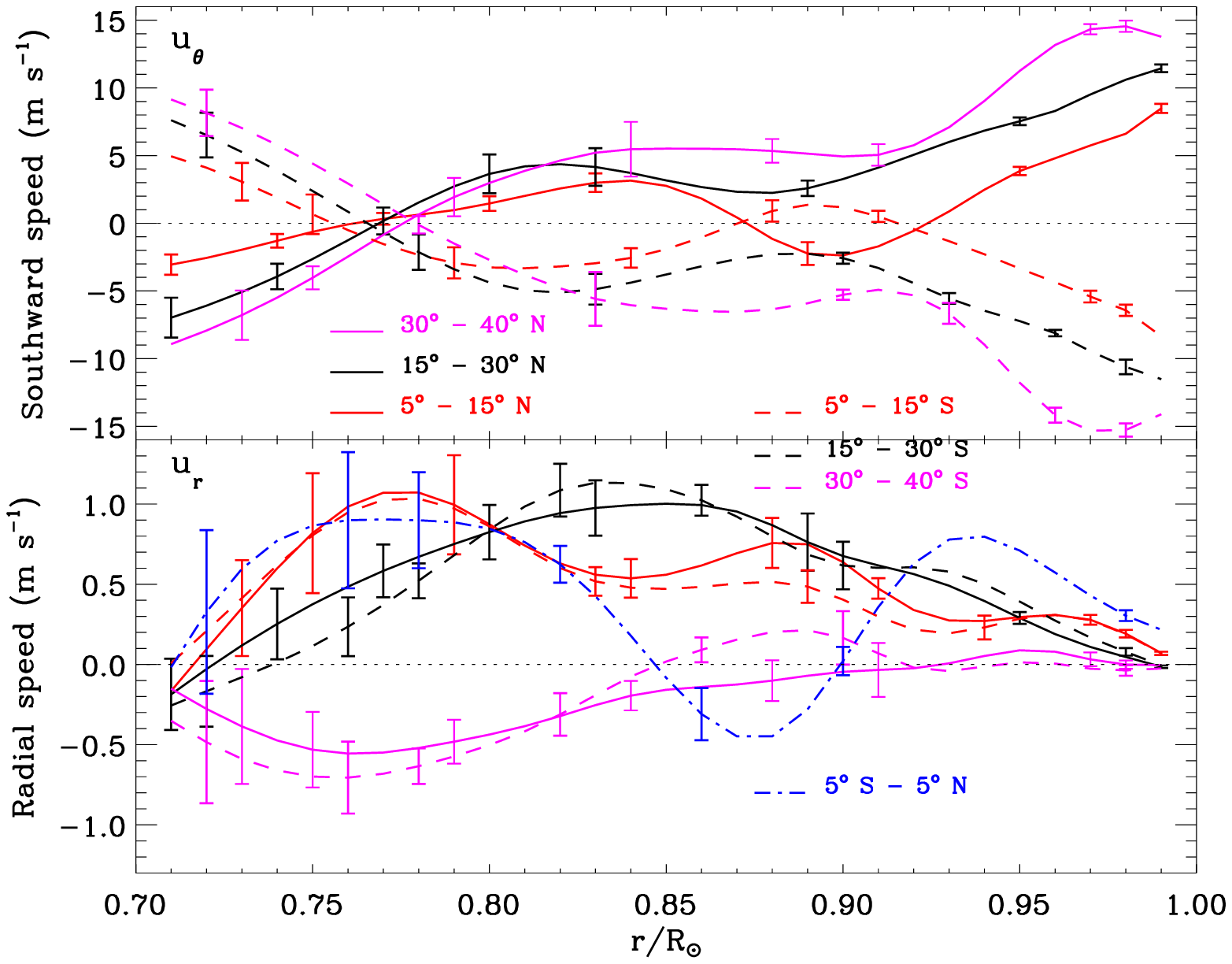}
\caption{Same as Figure 3 but with a lower smoothing
($\lambda_r=2\times10^{-4}, \lambda_\theta=2\times10^{-3}$).}
\label{fig4}
\end{figure*}

Further, the higher-latitude poleward flows seen down to depths of about 70 Mm (10\% of the solar radius from the surface) seem to 
drive deep downward flows beyond $40^{\circ}$ latitudes (refer to the right panel showing $u_{r}$ in Figure 3). This downward part of 
MC leads to the consistent signals of equator-ward return flow seen below about $0.77R_{\odot}$ over the whole range of latitudes 
(left panel of Figure 4). Upwards flows at low latitudes and downward flows at higher latitudes, as captured in the inverted $u_{r}$ 
(right panels of Figure 3 and 4), peak at a depth of about $0.79R_\odot$ with amplitudes of about 1 m s$^{-1}$.

\subsection{Tests with Artificial Data}

In order to check if with the current level of errors in the travel-times it is
possible to distinguish between the solution with single cell and that with
a double cell, we generated artificial data with some assumed meridional
flow profiles and added random errors to these which are consistent with
those in observed data. Using these artificial data we did the inversions
using the same (higher values) smoothing parameters to check for reliability of inversion
results. Here we present results from two different meridional flow profiles,
one a single cell extending till $0.7R_\odot$ and another double cell profile
with first cell ending at $0.80R_\odot$. In both cases the surface velocity is
chosen to be close to solar value to make meaningful comparison.
The results are shown in Figures 5. 
\begin{figure*}[ht]
\epsscale{1.16}
\centering
%\figurenum{2}
% I believe the dotted lines are true values?
%\plottwo{uth1_in_vs_out_r_artflows_nn.eps}{ur1_in_vs_out_r_artflows_nn.eps}
%\plottwo{uth2_in_vs_out_r_artflows_nn.eps}{ur2_in_vs_out_r_artflows_nn.eps}
\plottwo{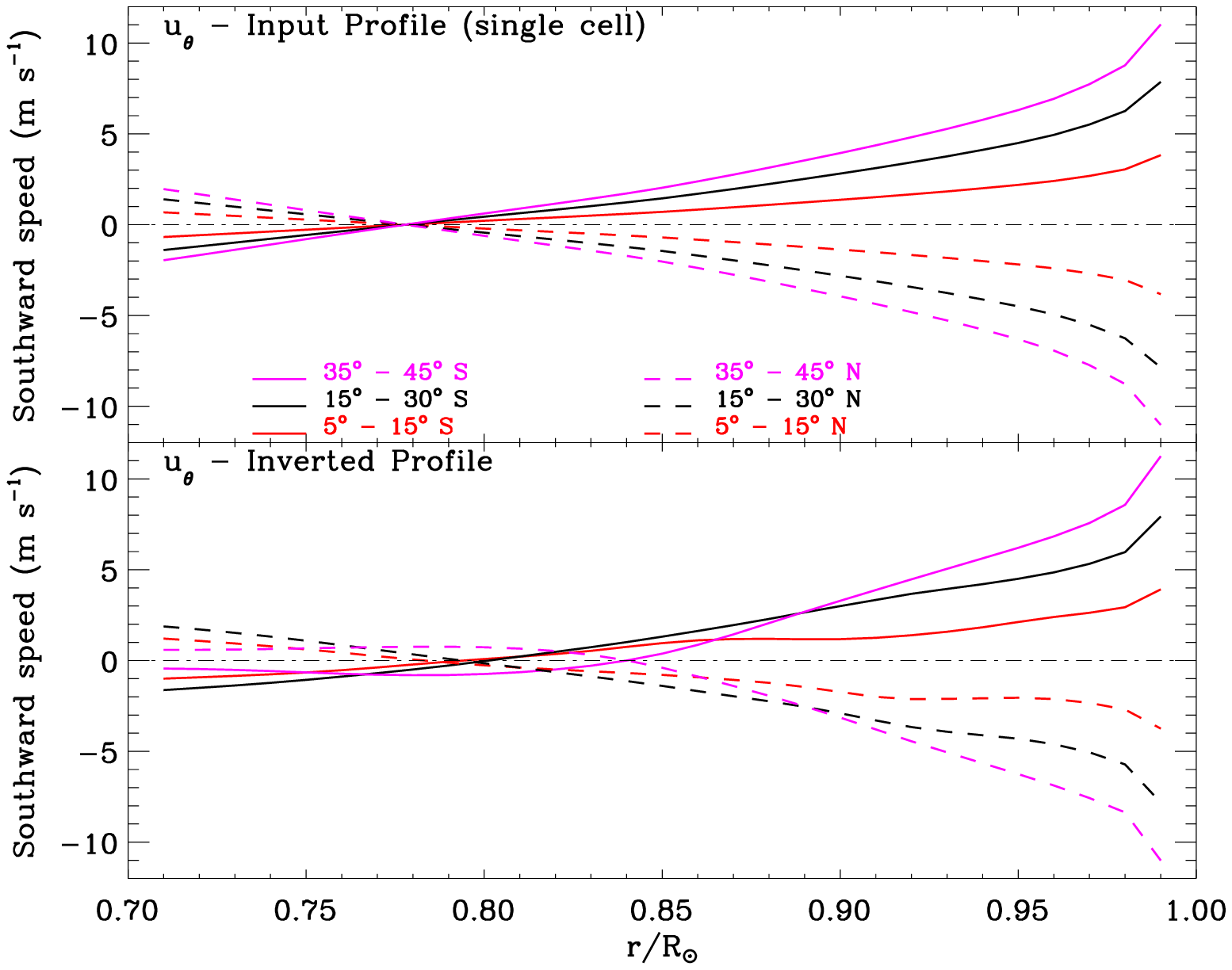}{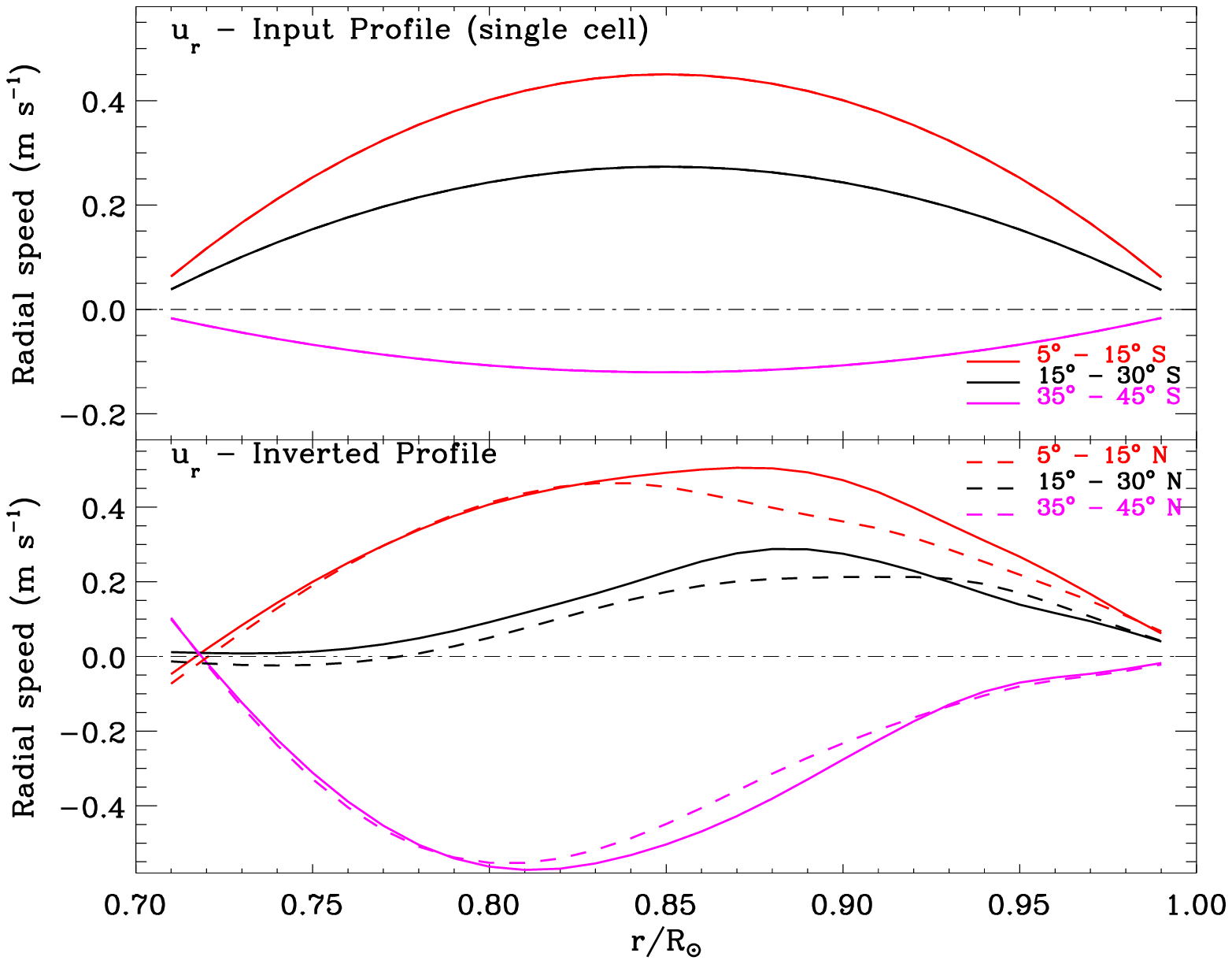}
\plottwo{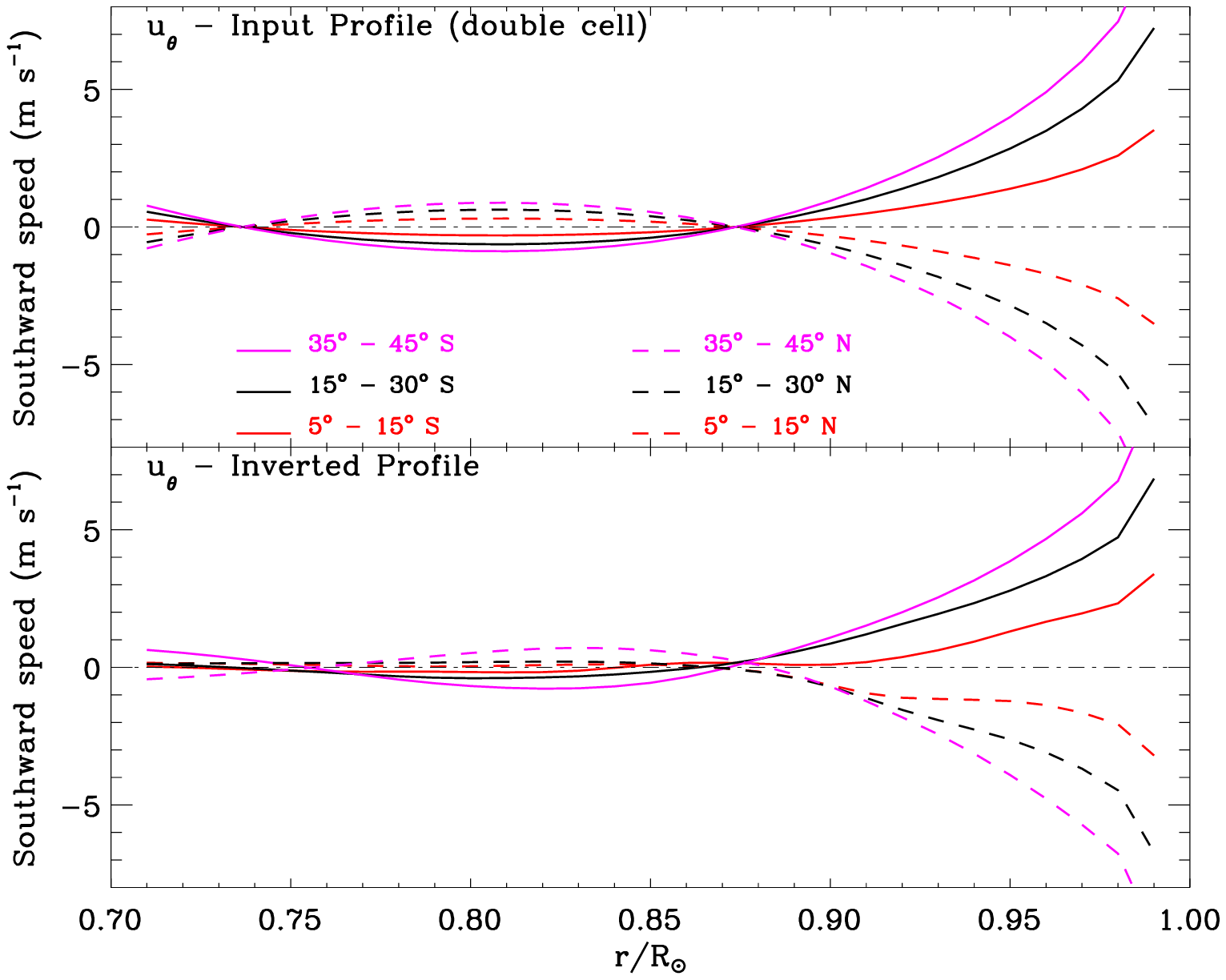}{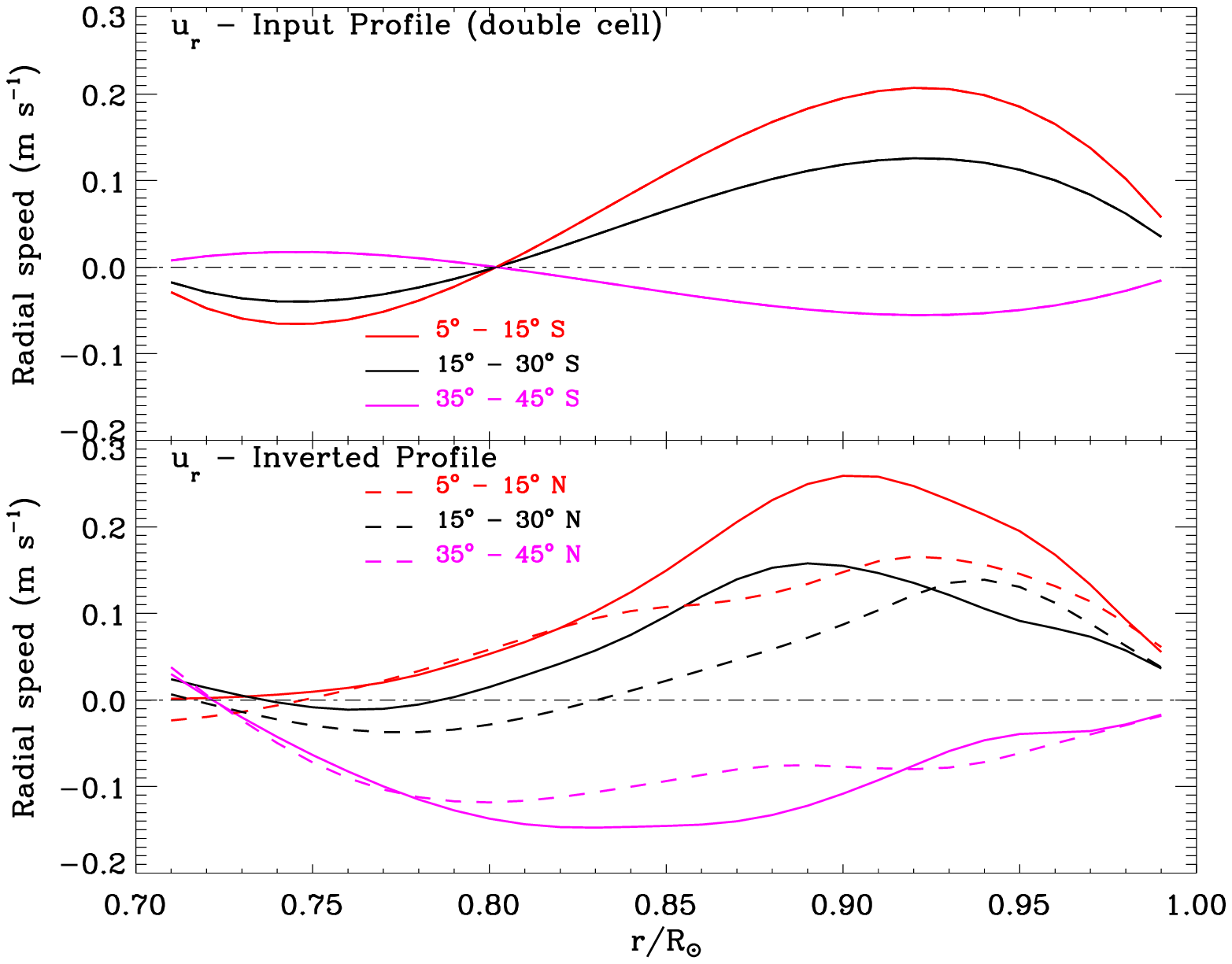}
\caption{Cuts across depth of artificial data test inversions for $u_{\theta}$ and $u_{r}$ averaged over three different latitude 
ranges, as marked in the panel The upper part of the figure shows the meridional velocity components for artificial data for MC with 
single cell, while the lower part shows that for MC with double cell. The upper panels in each part show the input MC profile while 
the lower panels show the results obtained by inversion of artificial data. The errorbars are not shown in lower panel, but the errors 
are same as those shown in Figure 3 for observed data.}
\label{fig5}
\end{figure*}
 
The results for single cell profile are shown in the top row of Fig.~\ref{fig5} which compares the flow velocity in the input profile 
with those obtained by inversion of travel times. It can be seen that the agreement is fairly good and in most region the difference 
between the true value and inverted value of $u_\theta$ is less than 1 m s$^{-1}$. The sign change in $u_\theta$ is also seen in the 
inverted profiles, though at a slightly deeper layer. Similarly, the bottom row of Fig.~\ref{fig5} shows the results for double cell 
profile of MC. In this case the agreement is even better than that for single cell profile. The inverted profiles clearly show two 
sign changes in $u_\theta$ with radius. Thus it is clear that our inversions are able to distinguish between the single and double 
cell profiles even in presence of observational errors. It may be noted that in both these cases the velocity in lower layers 
($r<0.9R_\odot$) is rather small and still inversions can detect the signal reliably. The radial component of velocity, $u_r$ is very 
small for double cell pattern and it is difficult to determine it reliably in the deep interior, where it is of order of 0.1 m 
s$^{-1}$.

These test results give some confidence in our results for the Solar MC
and it is likely that solar MC is a single cell circulation pattern
similar to what is assumed in flux transport dynamo models.\\

%\clearpage
\section{Discussion and Conclusion} \label{concl}

Currently, there is an increased focus on helioseismic studies of the MC in the deep layers of the convection zone owing to several 
recent developments. Firstly, recent years have seen intense developments in modeling the solar dynamo owing to several curious 
features of the last few solar cycles, including the extended minimum between cycles 23 and 24 and the overall gradual decline of the 
strengths of the past few cycles (see \citet{pcharb14} and references therein). Secondly, almost two solar cycles long continuous 
helioseismic data is now available from GONG, MDI/SOHO and the new HMI/SDO, which can help in addressing solar interior variations 
related to the solar dynamo and the cycle. Temporal variations of large-scale motions in the interior of the Sun, {\it e.g.,} the solar 
rotation and MC, can now be studied using these data. As far as the MC is concerned, measuring it over depth below about 10\% of the 
solar radius from the surface is a difficult proposition as the expected signals in helioseismic observables are very small and are 
plagued by noise and systematics.
%(it is to be noted that global helioseismic frequencies
%are insensitive to MC within the first order perturbations to them, and the second order changes are too small to measure successfully).
As described in Section 1, local helioseismic methods, ring diagram and time-distance helioseismology, have successfully mapped the 
near-surface part of the MC. The identification and empirical correction (or removal) for the large systematics, CLS, in helioseismic 
travel times by \citet{zhaoetal12,zhaoetal13} have now made studying the MC throughout the convection zone possible albeit with 
contentious inferences on the deep structure of MC \citep{zhaoetal13,jackiewiczetal15}.

Our work presented here provides an independent attempt to study MC in deep interior. We have made some improvements via adding 
additional physical constraints in the analysis procedures. As pointed out by \citet{jackiewiczetal15}, the existing results on the 
deep MC \citep{zhaoetal13,jackiewiczetal15} are problematic in satisfying the mass-conservation constraints. The depth variation of 
inverted flow velocity amplitudes (refer to Figure 4 of \citet{jackiewiczetal15}) run counter to what is expected from 
mass-conservation. We believe that our analysis which addresses mass-conservation in deriving both the horizontal and radial flow 
components, $u_{\theta}$ and $u_{r}$, simultaneously from the inversion of measured travel times has added a significant improvement 
in the time-distance helioseismic probing of the deep MC. Within the limits of errors in the data and in the inversion procedure, we 
find that the MC has possibly a large-scale return flow at depths below about $0.77R_{\odot}$. We do not see a shallow return flow at 
about $0.9R_\odot$ consistently throughout the whole latitude range covered ($\pm 60^{\circ}$), although, there are indications of 
sign reversals at latitudes less than about 30$^{\circ}$, but not above our error limits, in both $u_{\theta}$ and $u_{r}$ at this 
depth. We find that the inverted $u_{r}$ exhibits broad upwellings extending over a depth range of $0.75R_{\odot}$ -- $0.85R_{\odot}$ 
within about $\pm 15^{\circ}$ latitudes with a peak amplitude of about 1 m s$^{-1}$. At latitudes in the range of $30^{\circ}$ -- 
$40^{\circ}$ and beyond we find that the poleward flows with typical amplitudes of about 3 -- 5 m s$^{-1}$ over the depth range of 
$0.9R_{\odot}$ -- $0.8R_{\odot}$. This region is bounded by the deep downward flows beyond $40^{\circ}$ latitudes and forming a deep 
return flow at depths below $0.77R_{\odot}$. Among the theoretical models or predictions in the literature, the
MC profile calculated by \citet{kitchatinovolemskoy11} from a mean-field hydrodynamical model appear to be close to our findings here.
In particular, the sharp decline in $u_\theta$ observed over the depth range $0.97 - 0.9R_\odot$ matches their model as well as the transition to equatorward flow 
seen beneath the depth of $0.77R_{\odot}$.
Numerical simulations of \citet{feathmiesch15} show multiple cells of MC at low latitudes. Our inversion results also show some 
change in sign at low latitude, though the value is comparable to errorbars and is in a restricted range of latitude and depth.

To test if we could indeed detect multiple cells in MC we tried an exercise with artificial travel-time data generated using modeled
single- and double-cell MC profiles. We find that it is indeed possible to detect multiple cells in $u_\theta$, even though the velocity 
in deep layers is comparable to errorbars. These results tend to suggest that the solar MC probably consists of a single-cell pattern 
covering the entire convection zone, although, significant improvements in analysis are required to confirm this result.

\acknowledgments

We thank Junwei Zhao (Stanford University) for helping us with the data preparation at the Stanford JSOC 
Helioseismology pipeline. The HMI data used are courtesy of NASA/SDO and the HMI science team. Data intensive computations performed 
in this work were carried out using the High-Performance Compute Cluster of the Indian Institute of Astrophysics, Bangalore.

\end{document}